\documentclass{desyproc}

\begin{document}
\title{Core-Collapse Supernovae: Explosion Dynamics, Neutrinos and Gravitational Waves}

\author{{\slshape Bernhard M\"uller$^1$, Hans-Thomas Janka$^1$, Andreas Marek$^1$, Florian Hanke$^1$, Annop Wongwathanarat$^1$, Ewald M\"uller$^1$}\\[1ex]
$^1$Max-Planck-Institut f\"ur Astrophysik, Karl-Schwarzschild-Str. 1, 85748 Garching, Germany}

\contribID{xy}

\confID{1964}  
\desyproc{DESY-PROC-2010-01}
\acronym{PLHC2010} 
\doi  

\maketitle

\begin{abstract}
The quest for the supernova explosion mechanism has been one of the
outstanding challenges in computational astrophysics for several
decades. Simulations have now progressed to a stage at which the
solution appears close and neutrino and gravitational wave signals
from self-consistent explosion models are becoming available.  Here we
focus one of the recent advances in supernova modeling, the inclusion
of general relativity in multi-dimensional neutrino hydrodynamics
simulations, and present the latest simulation results for an
$11.2 M_\odot$ and a $15 M_\odot$ progenitor. We also mention
3D effects as another aspect in supernova physics awaiting
further, more thorough investigation.
\end{abstract}

\section{Introduction}
Massive stars end their lives as a core-collapse supernova (SN), a
violent event that involves the collapse of the iron core of the
progenitor to a proto-neutron star and the subsequent expulsion of the
outer layers of the star with a kinetic energy on the order of
$10^{51} \ \mathrm{erg}$, which is associated with a spectacularly
bright optical display. Currently, there is still no final consensus
on the supernova explosion mechanism that operates in the optically
obscured supernova core, and a number of competing ideas are
under discussion.  The delayed neutrino-driven mechanism
\cite{wilson_85,bethe_85}, which relies on neutrino energy deposition
in the gain region to revive the stalled shock, remains the most
promising candidate, provided that the efficiency of neutrino heating
can be sufficiently enhanced by multi-dimensional hydrodynamical
instabilities such as convection and the so-called standing accretion
shock instability SASI \cite{blondin_03,foglizzo_06}. This mechanism
has worked successfully in several recent 2D simulations
\cite{buras_06_b,marek_09,bruenn_10} (some of which 
appeared to be only marginally successful \cite{marek_09}), but has failed in others
\cite{burrows_06,burrows_07}. Alternatives to the neutrino-driven
mechanism have also been proposed, such as the acoustic mechanism
\cite{burrows_06,burrows_07} (whose viability has been called into
question by \cite{weinberg_08}, however), magnetohydrodynamically
driven supernovae \cite{akiyama_03,dessart_07_a}, and explosions
triggered by a QCD phase transition \cite{sagert_09}.

As the ``engine'' driving this explosion is not directly accessible by
classical, photon-based astronomical observations, our understanding
of the supernova explosion mechanism has largely rested on numerical
simulations in the past ever since the pioneering work of
\cite{colgate_66}. Over the years, a variety of ambitious numerical
approaches has been developed to cope with the challenging interplay
of neutrino transport, multidimensional hydrodynamics, general
relativity (GR), neutrino physics, and nuclear physics in the supernova
problem. The currently most advanced models rely on sophisticated
multi-group neutrino transport schemes (e.g. ray-by-ray variable
Eddington factor transport \cite{rampp_02, buras_06_a},
ray-by-ray-diffusion \cite{bruenn_10}, 2D multi-angle transport
without energy bin coupling \cite{ott_08_a}, or the isotropic
diffusion source approximation \cite{liebendoerfer_09}) with different
strengths and weaknesses, and there have only been very tentative
attempts to venture forth to 3D models with these methods
\cite{bruenn_10,takiwaki_11}. Elements missing or only partially
included in current state-of-the-art-models, such as 3D effects (whose
potential is presently being debated
\cite{nordhaus_10,takiwaki_11,hanke_11}) or general relativity may
hold the key to a better understanding of the explosion
mechanism. Moreover, an improved treatment of such as yet poorly
explored aspects is also indispensable for accurate predictions of the
neutrino and gravitational wave signal -- the only \emph{observables} that
directly probe the dynamics in the supernova core.

Among the aspects that have not yet been thoroughly investigated in
self-consistent multi-D neutrino hydrodynamics simulations of
core-collapse supernovae, our group has recently begun to study the
influence of GR in more detail. Although the importance of
relativistic effects in core-collapse supernovae (due to the
compactness of the proto-neutron star and the occurrence of high
velocities) has long been recognized an demonstrated \cite{bruenn_01}, the
combination of GR hydrodynamics and multi-group neutrino transport has
long been feasible only in spherical symmetry
\cite{bruenn_01,yamada_99,liebendoerfer_04}.  With the relativistic
generalization of the ray-by-ray variable Eddington factor method
\cite{mueller_10} used in our neutrino hydrodynamics code
\textsc{Vertex}, we are now able to present first results about the
impact of GR on the explosion dynamics and, in particular, the
neutrino and gravitational wave emission in axisymmetric (2D)
supernova models.

\section{General Relativistic Effects in Multi-Dimensional
Supernova Models}

\begin{figure}[tb]
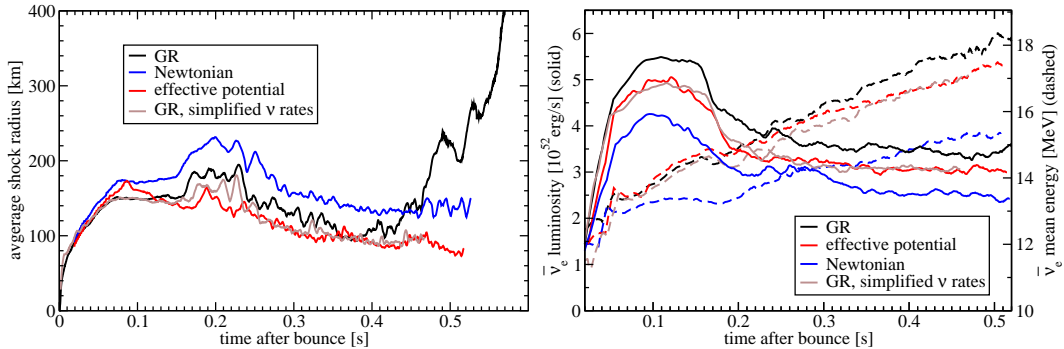

\centerline{
  \includegraphics[width=0.48\textwidth]{f1a.eps}
  \includegraphics[width=0.48\textwidth]{f1b.eps}}
\caption{Left: Average shock radius for the $15 M_\odot$ model as
  obtained in GR with our best set of opacities (black curve) and with
  simplified neutrino rates (light brown curve), in the purely
  Newtonian approximation (blue), and with an effective gravitational
  potential (red). Right: Electron antineutrino luminosities (solid)
  and mean energies (dashed) at the gain radius for these three cases.
\label{fig:1}
}
\end{figure}

\begin{figure}[tb]
\centerline{
  \includegraphics[width=0.48\textwidth]{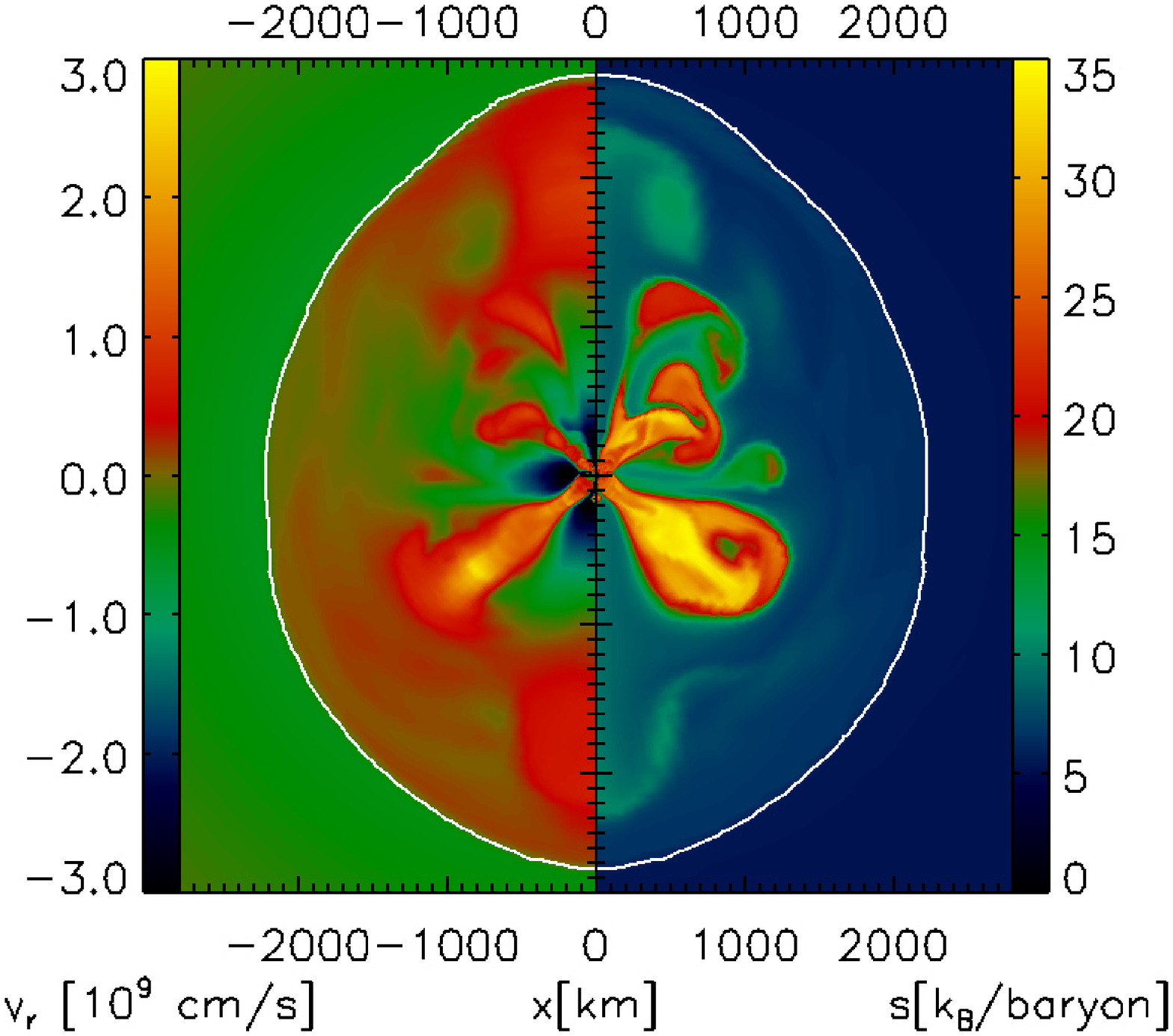}
  \includegraphics[width=0.48\textwidth]{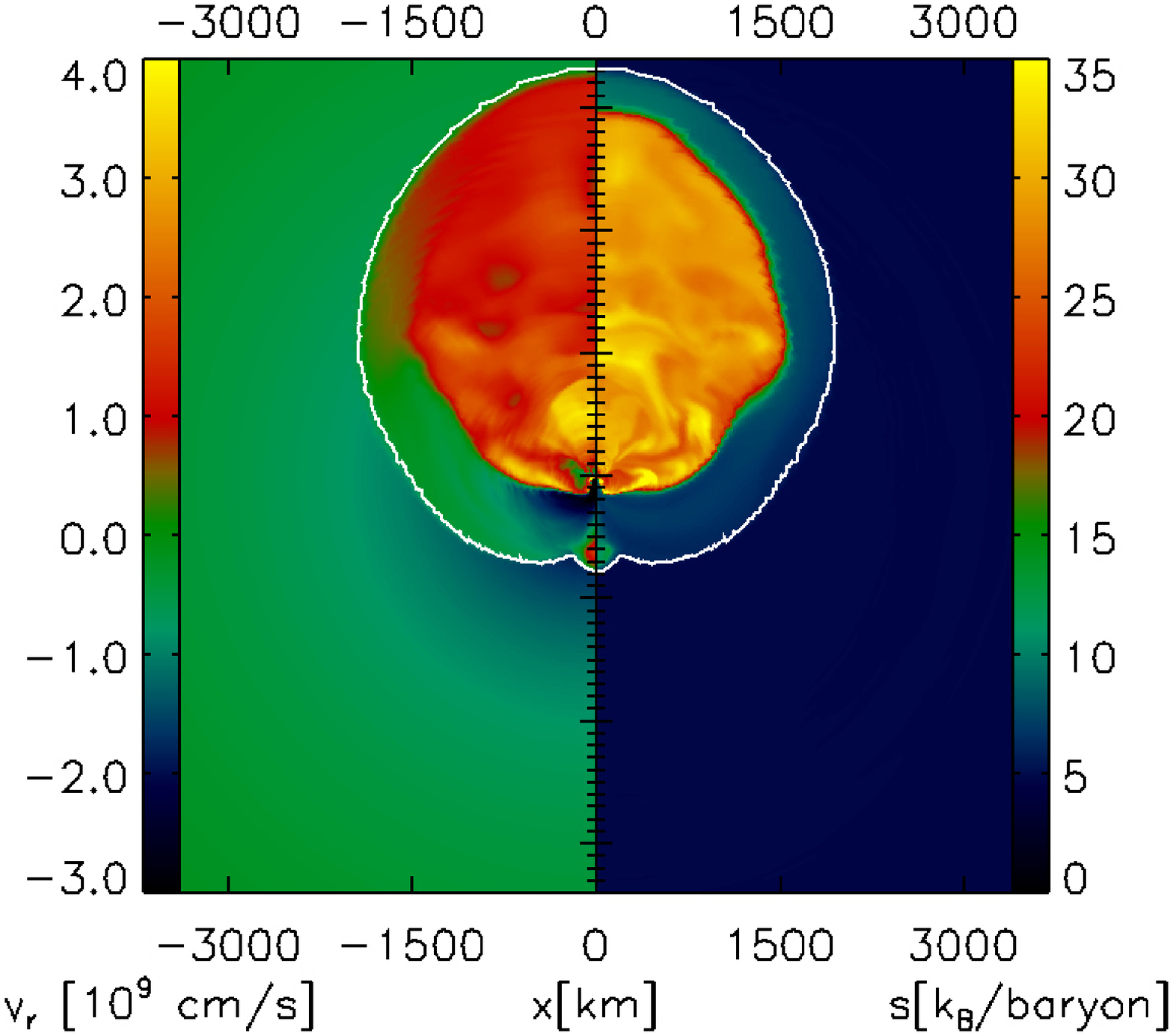}}
\caption{Explosion geometry for the GR simulations of the $11.2
  M_\odot$ (left panel, almost spherical shock) and the $15 M_\odot$
  progenitor (right panel, strong dipolar shock deformation) $658
  \ \mathrm{ms}$ and $745 \ \mathrm{ms}$ after bounce,
  respectively. The left and right half of the panels show the
  electron fraction $Y_e$ and the entropy $s$, respectively, and the
  shock is indicated as a white curve.
\label{fig:2}}
\end{figure}

\begin{figure}[tb]
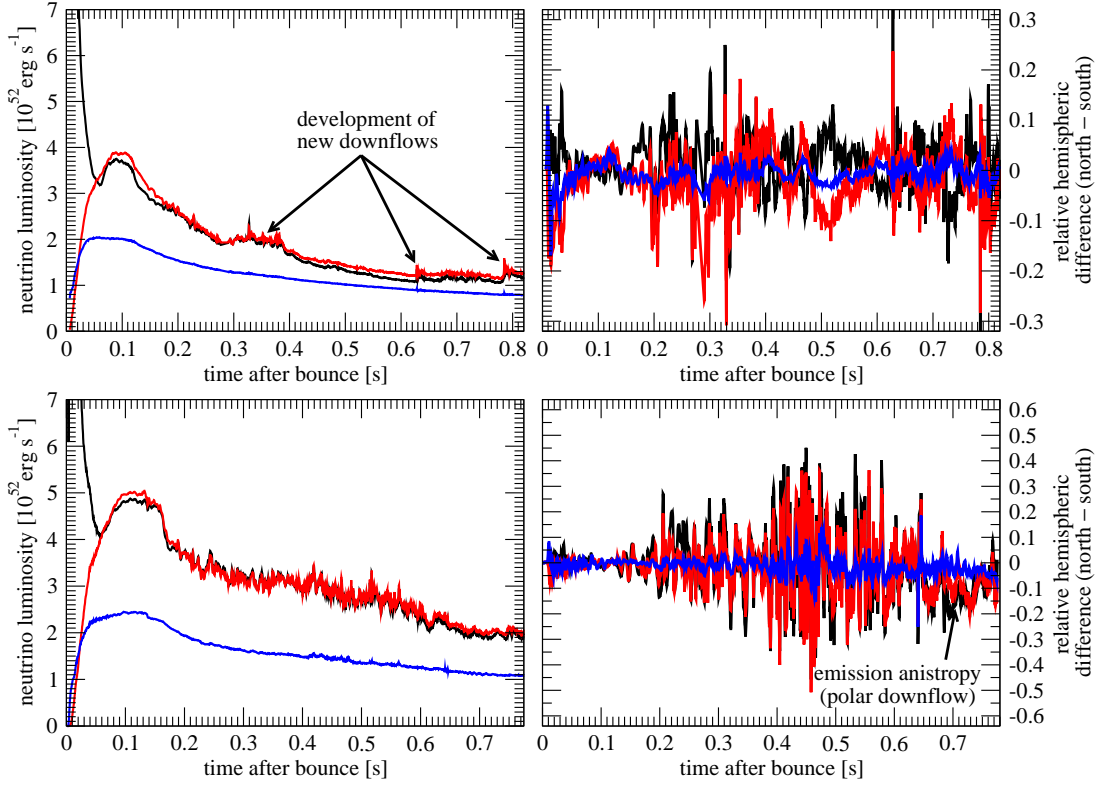

\centerline{
  \includegraphics[width=\textwidth]{f3a.eps}}
\centerline{
  \includegraphics[width=\textwidth]{f3b.eps}}
\caption{Left panels: Neutrino luminosities (defined as the total
  angle-integrated neutrino flux from the supernova) for the
  relativistic $11.2 M_\odot$ (top) and $15 M_\odot$ (bottom) models.
  Right panels: Relative differences of the angle-integrated neutrino
  fluxes $L_\mathrm{north}$ and $L_\mathrm{south}$ in the northern and
  southern hemisphere, computed as $2
  (L_\mathrm{north}-L_\mathrm{south}) /
  (L_\mathrm{north}+L_\mathrm{south})$.  Black, red, and blue curves
  are used for $\nu_e$, $\bar{\nu}_e$, and $\nu_{\mu/\tau}$,
  respectively.
\label{fig:3}}
\end{figure}

Our group has recently conducted relativistic supernova simulations
for progenitors with $11.2 M_\odot$ \cite{woosley_02} and $15 M_\odot$
\cite{woosley_95} well into the explosion phase, which were
supplemented by three additional runs for the $15 M_\odot$ star. In
order to estimate the magnitude of GR effects, two complementary
models were computed using either the purely Newtonian approximation
or the ``effective potential'' approach \cite{marek_06},
which has long been the only means of including some GR corrections in
multi-D neutrino hydrodynamics simulations. In addition, we also
calculated a model with a simplified set of neutrino opacities
(neglecting the effects of recoil, high-density correlations and weak
magnetism in neutrino-nucleon reactions and ignoring reactions between
different neutrino flavors), which serves to illustrate the
importance of the neutrino microphysics for the dynamics in the
supernova core and provides a scale of reference for the GR effects.
As a marginal case close to the threshold between explosion and
failure \cite{marek_09}, the $15 M_\odot$ progenitor is ideally
suited for such a comparative analysis.

Interestingly, we find that among the $15 M_\odot$ models, the GR run
with improved rates is the only one to develop an explosion with shock
revival occurring some $450 \ \mathrm{ms}$ after bounce
(Fig.~\ref{fig:1}), indicating the relevance of both GR effects
\emph{and} of the neutrino microphysics. The different evolution of
the three models with a different treatment of gravity is a
consequence of the different compactness and surface temperature of
the proto-neutron star, which leads to a clear hierarchy of the electron
neutrino and antineutrino luminosities and mean energies (which
determine the heating conditions) between the three cases
(cp. \cite{bruenn_01,liebendoerfer_05,mueller_10} for this effect in
1D simulations) as illustrated by Fig.~\ref{fig:1} for the electron
antineutrinos, where the enhancement in GR is most pronounced. The
beneficial effect of higher local heating rates as compared to the
Newtonian case is, however, counterbalanced by the faster advection of
material through the gain layer around a more compact proto-neutron
star in the effective potential run, but in the GR case, the enhanced
heating is strong enough to overcome this adverse effect.

It is noteworthy that the neutrino emission and the shock evolution
are similarly sensitive to the neutrino interaction rates (in agreement
with the findings of \cite{rampp_proc_02,bruenn_10}). In the run
with improved microphysics, weak magnetism and nucleon correlations
lower the opacities for $\bar{\nu}_e$, shift the neutrinosphere into
deeper and hotter regions of the proto-neutron star surface
\cite{horowitz_02}, and thus result in harder $\bar{\nu}_e$ spectra
(by up to $\sim 1 \ \mathrm{MeV}$ during the late phases) and
increased $\bar{\nu}_e$ luminosities, which also allows for more
efficient heating in the gain layer. Neglecting possible effects of
flavor conversion and MSW, this would also imply somewhat higher
detection rates for $\bar{\nu}_e$ (by $\sim 15 \%$). 

\section{Neutrino and Gravitational Wave Signals from Supernovae}
Both the $11.2 M_\odot$ and the $15 M_\odot$ models have been evolved
well into the post-explosion phase until $\sim 0.8 \ \mathrm{s}$ after
bounce, and thus provide a good illustration of the impact of
multi-dimensional effects on the neutrino and gravitational wave
signal during the different stages of the evolution. Prior to the
onset of the explosion the neutrino luminosities of both
models (Fig.\ref{fig:3}, left panels) are characterized by the familiar
large contribution of the accretion luminosity for $\nu_e$ and
$\bar{\nu}_e$. As soon as the SASI starts to grow vigorously, we also
observe the strong angle-dependent time variations in the neutrino
flux (particularly in $\nu_e$ and $\bar{\nu}_e$ , see
Fig.~\ref{fig:3}, right panels) that have been discussed in
\cite{ott_08_a,marek_08,lund_10,brandt_10} and can potentially be used
to extract the frequencies of the SASI from the neutrino signal using
detectors with high temporal resolution such as Icecube
\cite{lund_10}. In our simulations, these fluctuations are present in
similar strength as in Newtonian and effective potential models
\cite{ott_08_a,marek_08,brandt_10} with the hemispheric flux
fluctuating by several tens of percent; and the dominant frequencies
($45 \ \mathrm{Hz}$ and $75 \ \mathrm{Hz}$ for the $\ell=1$ and
$\ell=2$ mode) are in excellent agreement with
\cite{marek_08,lund_10}.

It is interesting to note that our results suggest that the neutrino
signal changes its character only gradually over several hundreds of
milliseconds after the onset of the explosion.  The temporal
fluctuations are actually strongest during the first $\sim 200
\ \mathrm{ms}$ after shock revival, and the luminosities of $\nu_e$
and $\bar{\nu}_e$ do not show any abrupt decline correlated with the
time of the explosion. This behavior is in marked contrast to the
abrupt drop in the $\nu_e$ and $\bar{\nu}_e$ luminosities in 1D models
with artificial explosions \cite{fischer_10}, and is due to the fact
that quite large accretion rates can still be maintained through the
downflows at late times in multi-dimensional models
(Fig.~\ref{fig:2}). As long as the shock does not expand too rapidly,
new downflows may still form and channel fresh material into the
cooling region (see \cite{marek_09}, as in the case of the $11.2
M_\odot$ progenitor, which leads to noticeable ``bumps'' in the
neutrino luminosity (Fig.~\ref{fig:3}, top left panel).  For the $15
M_\odot$ progenitor, an even higher accretion luminosity can be
maintained continuously because of the presence of a stable polar
downflow in an extremely asymmetric explosion geometry
(Fig.~\ref{fig:2}). With a single downflow, the neutrino emission
exhibits a strong directional dependence with sustained hemispheric
flux differences of up to several tens of percent
(Fig.~\ref{fig:3}). On the other hand, the strong high-frequency
fluctuations subside during the late phases as the expansion of the
shock quenches further SASI activity. The neutrino signal thus still
reflects the dynamical evolution of explosion models in multi-D,
albeit in a form very different from artificial 1D explosions.

\begin{figure}[tb]
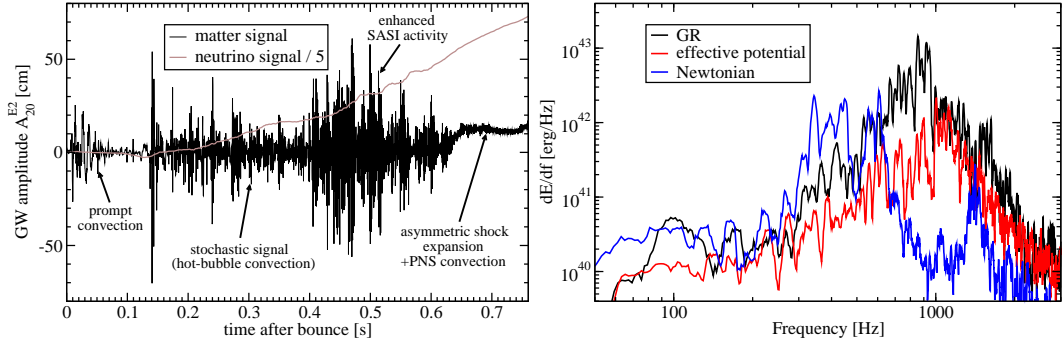

\centerline{
  \includegraphics[width=0.48 \textwidth]{f4a.eps}
  \includegraphics[width=0.48 \textwidth]{f4b.eps}}
\caption{Left: Matter (black) and neutrino (light brown) gravitational
  wave signals for the general relativistic $15 M_\odot$ explosion
  model. Right: Influence of the treatment of gravity (black: GR
  hydro, red: Newtonian hydro + effective potential, blue: purely
  Newtonian) on the gravitational wave energy spectrum for the first
  $500 \ \mathrm{ms}$ of the post-bounce evolution of the $15 M_\odot$
  progenitor.
\label{fig:4}}
\end{figure}

Naturally, the determination of gravitational wave signals has also
been among the major goals of the simulations with the relativistic
version of \textsc{Vertex}. Qualitatively, we obtain similar waveforms
as computed in the Newtonian or effective potential approximation
\cite{marek_08,murphy_09,yakunin_10} with clearly distinct phases in
the signal corresponding to the dynamics (Fig.~\ref{fig:4}, cp.
with ref. \cite{murphy_09}). Shortly
after bounce, prompt convection and early SASI activity produce a
low-frequency, quasi-periodic signal, which is followed by a more quiescent
period until hot-bubble convection and strengthening
SASI sloshing motions
gives rise to a stochastic signal
with typical frequencies rising from $500 \ \mathrm{Hz}$ to over $1000
\ \mathrm{Hz}$ during a phase of $\sim 200 \ \mathrm{ms}$ around shock revival
revival, when gravitational wave emission is strongest. Afterwards,
proto-neutron star convection becomes the dominant source of
high-frequency gravitational waves. In the case of the $15 M_\odot$
progenitor with a rather extreme explosion geometry, asymmetric shock
expansion and neutrino emission also give rise to a monotonously rising
``tail signal'' which contributes somewhat to the low-frequency part
of the spectrum. Despite the qualitative similarities of waveforms in
GR and the Newtonian approximation, GR effects have a considerable
impact on the power spectrum, however. The integrated signal for the
first $500 \ \mathrm{s}$ (Fig.~\ref{fig:4}) peaks at considerably
higher frequencies in GR ($\sim 900 \ \mathrm{Hz}$) compared to the
purely Newtonian case ($\sim 500 \ \mathrm{Hz}$). On the other hand,
the effective potential approximation even overestimates the
peak frequency ($\sim 1100 \ \mathrm{Hz}$), because the lower
proto-neutron star surface temperature leads to a higher
Brunt-{V\"ais\"al\"a} frequency and therefore to a more abrupt braking
of convective bubbles at the lower boundary of the hot-bubble
convection region (cp. with the interpretation of the
characteristic frequencies given in \cite{murphy_09}).

\section{Outlook -- 3D Supernova Modeling}
The results presented here demonstrate that both GR effects and
variations in the neutrino microphysics have a significant impact on
the neutrino emission and, consequently, and on the dynamics in the
supernova core. As illustrated by the marginal $15 M_\odot$
progenitor, a detailed and sophisticated treatment of gravity and
neutrino interactions can very well be crucial for the success of the
neutrino-driven explosion mechanism. Such improvements in the models
also bring up the perspective of reliable, non-parametrized
predictions for the neutrino and gravitational wave signal
\emph{beyond} the accretion phase (cp. also \cite{yakunin_10}), whose
salient features have been pointed out in the last section.

One of the major limitations of the models discussed here is their
restriction to axisymmetry, which presently remains a necessary
compromise for simulations with the most advanced multi-group
neutrino transport methods. In the meantime, 3D effects can already be
explored with the help of parametrized approaches and cheaper
approximative methods to gain insights into their potentially
important role for the explosion mechanism
\cite{nordhaus_10,takiwaki_11,hanke_11} and the expected changes in
the neutrino and gravitational wave signals
\cite{wongwathanarat_11}. On the background of the strong sensitivity
of the heating conditions on the neutrino treatment, conclusions about
the implications of 3D effects for the viability of the
neutrino-driven mechanism can only be drawn with some caution,
however. Recent studies by \cite{takiwaki_11} and by our own group
\cite{hanke_11} have indeed demonstrated that models do not necessarily
explode more easily in 3D than in 2D \cite{takiwaki_11,hanke_11}, and
have rather pointed out issues that require further investigation,
such as the role of feedback effects of convection and the SASI on the
neutrino emission\cite{takiwaki_11,hanke_11}, dimensionality-dependent
resolution effects due to the different direction of the turbulent
cascade \cite{hanke_11}, and the growth and saturation of the SASI in
3D \cite{hanke_11}.

While the influence of the dimensionality on the explosion conditions
remains a controversial topic, the gravitational wave
\cite{scheidegger_08,kotake_09,scheidegger_10,kotake_11} and neutrino
signatures of non-radial hydrodynamic instabilities developing
during the post-bounce phase will undoubtedly be affected by going
from 2D to 3D. First predictions based on models with simplified
semi-parametrized neutrino transport have recently become available
\cite{wongwathanarat_11}, and suggest that the lack of a preferred
direction in 3D and weaker activity of the $l=1$ SASI sloshing mode
reduce both the gravitational wave amplitude and the fast temporal
variations of the neutrino signal by a factor of several. Even these
findings are still subject to uncertainties about the dynamics in the
supernova core (in particular concerning the behavior of the SASI in
3D), and in the end, accurate signal predictions will also require
self-consistent simulations with at least the same level of
sophistication as currently available in 2D.

\section*{Acknowledgments}
This work was supported by the Deutsche
Forschungsgemeinschaft through the Transregional Collaborative
Research Centers SFB/TR 27 ``Neutrinos and Beyond'' and SFB/TR 7
``Gravitational Wave Astronomy'' and the Cluster of Excellence EXC 153
``Origin and Structure of the Universe''
(http://www.universe-cluster.de). The simulations of the MPA
core-collapse group were performed on the IBM p690 of the Computer
Center Garching (RZG), the NEC SX-8 at the HLRS in Stuttgart (within
the \textsc{SuperN} project), the Juropa Cluster at the John von
Neumann Institute for Computing (NIC) in J\"ulich (partially through a
DECI-6 grant of the EU), and on the IBM p690 at Cineca in Italy through a DECI-5
grant of the DEISA initiative.

\section{Bibliography}
 

\begin{footnotesize}

\bibliographystyle{h-physrev5}
\bibliography{paper}

\end{footnotesize}


\end{document}